\documentstyle[psfig]{caosp}
\begin{document}
\pubyear{1998}
\volume{23}
\firstpage{1}
\hauthor{M. Hempel {\it et al.}}
\title{Dusty and dust-free A stars}
\author{Marc Hempel, Hartmut Holweger and Inga Kamp}
\institute{Institut f\"{u}r Theoretische Physik und Astrophysik, 
Universit\"{a}t Kiel}

\maketitle
\begin{abstract}
We present preliminary results of our search for circumstellar 
absorption features in the Ca~K lines based on high S/N observations obtained
with the ESO~CAT/CES system.
\end{abstract}
\section{Introduction}
\label{intr}
Main-sequence A stars have shallow surface convection zones, therefore their 
composition responds sensitively to any `contamination' by processes of
diffusion or accretion. For example, the metal deficiency of the
$\lambda$\,Bootis stars indicates accretion of depleted gas after separation
of gas and dust in the stellar environment. Yet not all A stars with
circumstellar (CS) matter show chemical anomalies indicative of 
accretion. A prominent example is $\beta$\,Pictoris.

However, in most cases nothing is known about their composition and the
presence of CS gas. Accurate surface abundances of A stars that are positive
or negative IRAS detections and a sensitive search for CS lines permit to trace
the signature of accretion differentially and with a high sensitivity.

\section{Results}
Table~1 provides some data of our programme stars. Spectral types and
rotational velocities are adapted from Holweger \& Rentzsch-Holm (1995) 
or, if not available, from the Bright Star Catalogue
(Hoffleit \& Jaschek 1982). A detailed spectrum synthesis will be carried out
to improve the $v\sin i$ values and to determine the calcium abundances. 
The parallaxes and the visual magnitudes have been obtained from 
the HIPPARCOS catalogue. The version of the UVBYBETA code (Moon \& Dworetsky 
1985) modified by Napiwotzki et al. (1993) was used to determine $T_{\rm eff}$ 
and $\log g$ from Str\"{o}mgren photometry (Hauck \& Mermilliod 1990).

Stars marked as `dusty' are IRAS sources whose infrared excesses are attributed
to circumstellar dust (Cheng et al. 1991, 1992). The other stars are negative
IRAS detections. The last column indicates preliminary detections of narrow 
absorption components in the Ca~K lines.

Figure~1 shows the projected rotational velocities of the program stars as a
function of effective temperature. Filled squares indicate stars with
circumstellar components in Ca~K.

\vspace{-2mm}
\begin{table}[hbtp]
\small
\begin{center}
\caption{}
\label{t1}
\begin{tabular}{lrrlrrccccc}
\hline\hline
HR  &  V    & Plx & Sp. Type &  v sin i  &  Teff &   log g    & dusty? & CS \\
    & [mag] &  [mas]&          &   [km/s]  &   [K] &   [cm/s$^2$]&  & lines?\\      
\hline 
1483 &  4.99 &  14.34 & A2IV     &    43  &    8930 &  3.90   &   yes  & no \\
1666 &  2.78 &  36.71 & A3III    &   179  &    8105 &  3.58   &   no   & no \\
2020 &  3.85 &  51.87 & A5V      &   132  &    8200 &  4.24   &   yes  & yes\\
2491 & -1.44 & 379.21 & A1V      &    13  &   10135 &  4.31   &   no   & no \\
2550 &  3.24 &  32.96 & A7IV     &   205  &    7525 &  3.48   &   no   & yes\\
2763 &  3.58 &  34.59 & A3V      &   154  &    8480 &  3.91   &   no   & no \\
3485 &  1.93 &  40.90 & A1V      &   160  &    9250 &  3.79   &   no   & yes\\
3615 &  4.00 &  26.24 & A2-A3IVm &    45  &    8300 &  4.13   &   yes  & no \\ 
3685 &  1.67 &  29.34 & A2IV     &   133  &    9090 &  3.08   &   no   & yes\\
3863 &  5.30 &  14.85 & A3IV     &    39  &    7965 &  4.07   &   yes  & no \\ 
4138 &  4.72 &  12.60 & A2m      &     7  &    8965 &  3.59   &   ?    & no \\
4534 &  2.14 &  90.16 & A3V      &   121  &    8625 &  4.21   &   yes  & no \\
4796 &  5.78 &  14.91 & A0V      &   130  &   10265 &  4.45   &   yes  & yes\\ 
4828 &  4.88 &  27.10 & A0V      &   166  &    9210 &  4.24   &   yes  & no \\
4881 &  6.14 &  16.62 & A0III    &   126  &    7440 &  3.96   &   ?    & yes\\
5028 &  2.75 &  55.64 & A2V      &    85  &    9395 &  4.10   &   no   & yes\\
5367 &  4.05 &  13.19 & A0IV     &   132  &   10245 &  3.71   &   yes  & yes\\ 
5531 &  2.75 &  42.25 & A3IV     &    84  &    8240 &  3.95   &   no   & no \\
6378 &  2.43 &  38.77 & A2V      &    26  &    8850 &  3.90   &   no   & no \\
6519 &  4.78 &   7.65 & A0V      &   272  &    9930 &  3.09   &   yes  & yes\\
6549 &  5.25 &  23.71 & A5IV-V   &    41  &    8190 &  4.25   &   yes  & no \\
6556 &  2.08 &  69.84 & A5III    &   219  &    7960 &  3.62   &   no   & yes\\
\hline\hline
\end{tabular}
\end{center}
\end{table}
\vspace{-2mm}
Obviously stars with detectable CS gas are found preferably among rapid
rotators. This strengthens the findings of Holweger \& Rentzsch-Holm (1995)
and their supposition that $v\sin i$ is the prime factor responsible for the
presence, or absence, of CS absorption in Ca~K'. Hauck et al. (preprint 1997)
provide additional evidence for this conclusion 
from an independent study of a different sample of $\lambda$ Bootis stars.

The lack of stars with CS lines and small $v\sin i$ values is interpreted
tentatively by Holweger \& Rentzsch-Holm (1995): for stars with 
CS gas concentrated in a disk-like structure the column density of absorbing
gas along the line of sight will be at its maximum if the disk is viewed
edge-on. Therefore CS lines should be detected preferably in objects with
$\sin i \approx 1$. Hence the chance to find 
a rapid rotator with CS lines in the lower part of the diagram is small.

Figure~2~ is an HR~diagram. The symbols are the same as in Figure~1. 
The absolute magnitudes have been calculated using the parallaxes found by
HIPPARCOS. The solid line represents the ZAMS (Vandenberg 1985). Stars with
CS gas are found both near and above the ZAMS.

\begin{figure}[hbt]
\hspace{1.5cm}
\psfig{figure=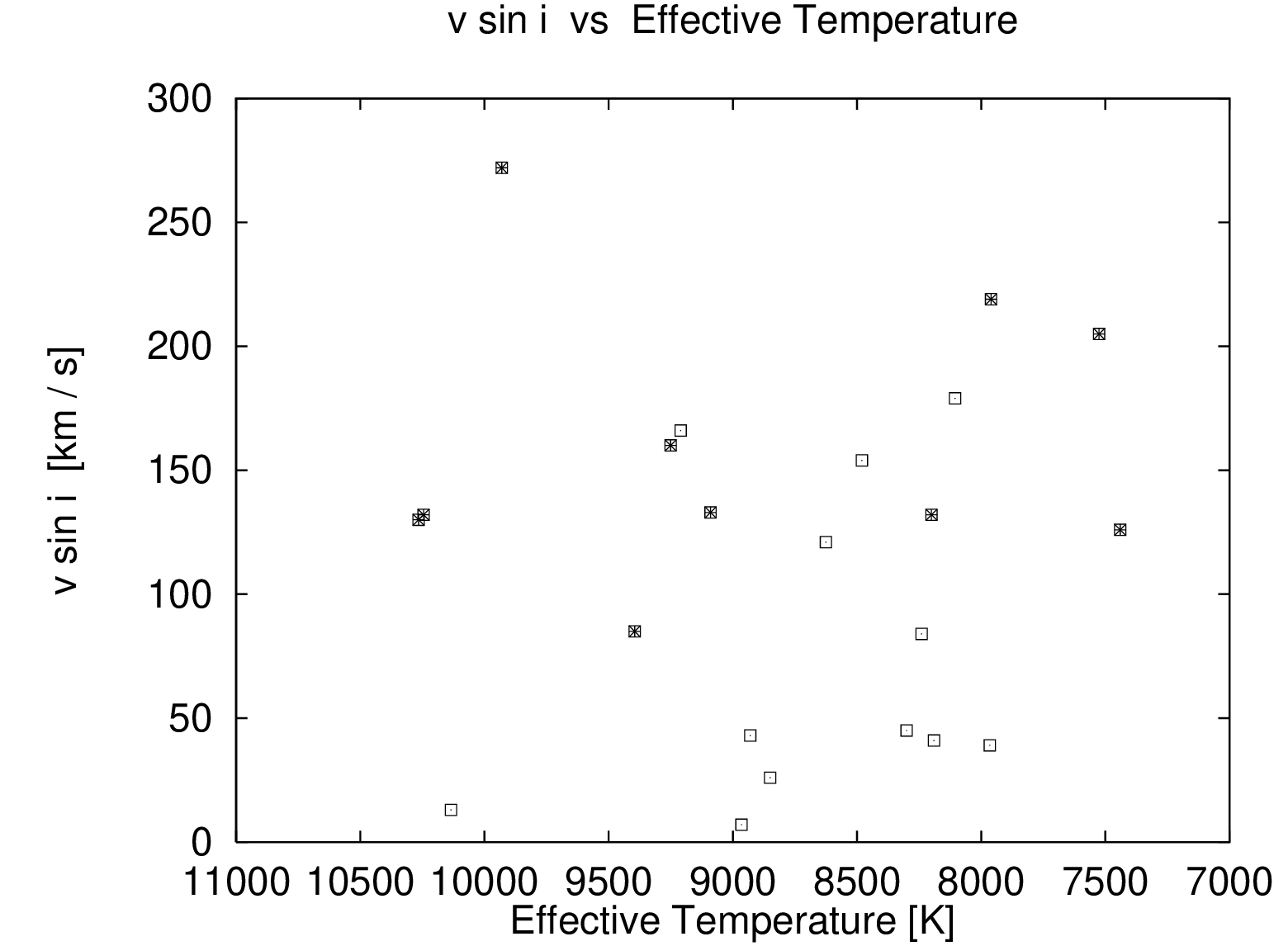,height=5.5cm}
\caption{}
\label{fp}
\end{figure}
\begin{figure}[hbt]
\hspace{1.5cm}
\psfig{figure=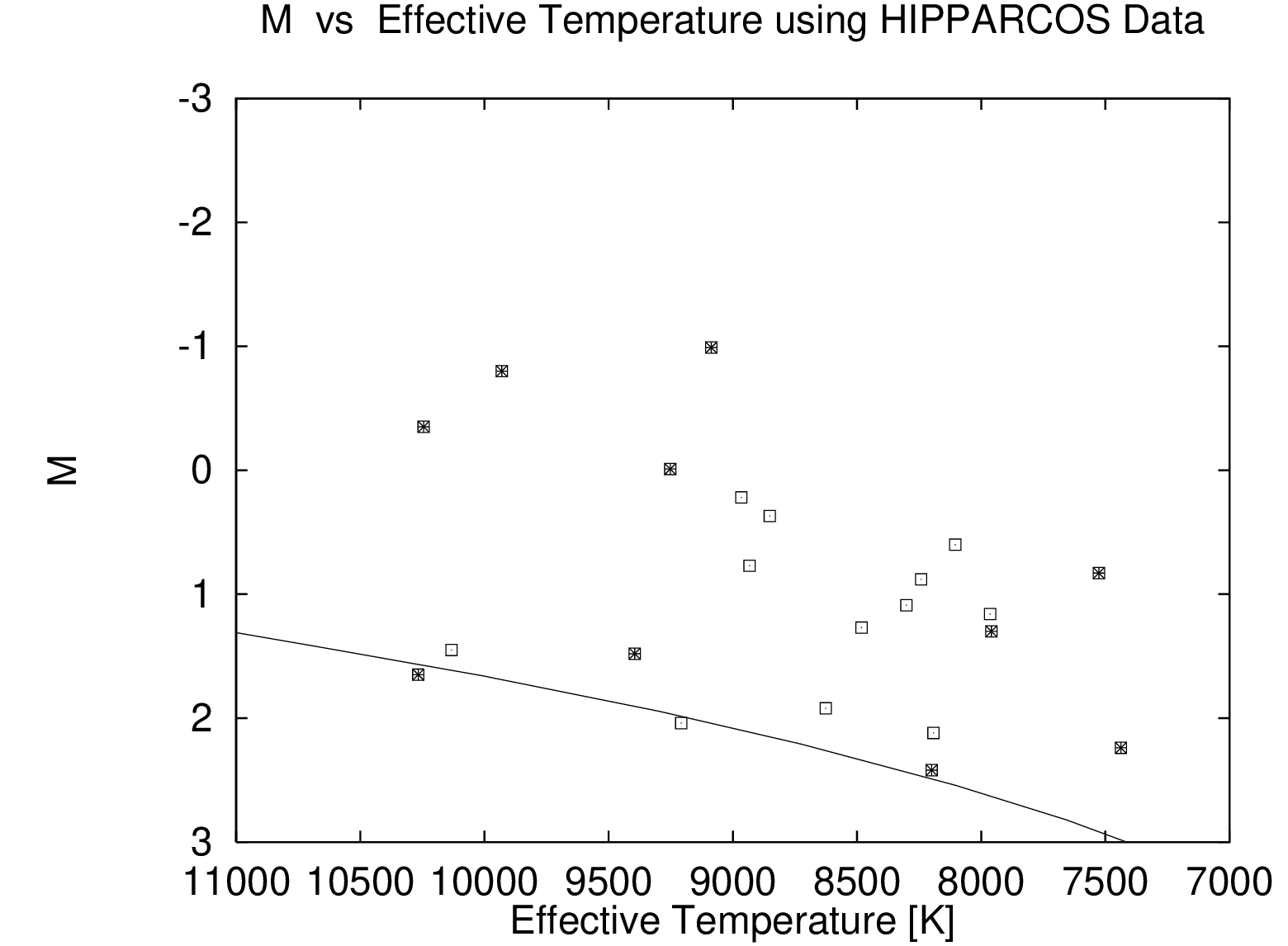,height=5.5cm}
\caption{}
\label{fp}
\end{figure}

\end{document}